\documentstyle[12pt]{article}
\begin{document}
\textheight 22cm
\textwidth 15cm
\topmargin 1mm
\oddsidemargin 5mm
\evensidemargin 5mm
\newcommand\extra{\vskip 10pt}
\newcommand{\bfl}{\begin{flushleft}}
\newcommand{\efl}{\end{flushleft}}
\newcommand{\bc}{\begin{center}}
\newcommand{\ec}{\end{center}}
\newcommand\ie {{\it i.e.}}
\newcommand\eg {{\it e.g.}}
\newcommand\etc{{\it etc.}}
\newcommand\cf {{\it cf.}}
\newcommand\viz{{\it viz.}}
\newcommand\grad{\nabla}
\newcommand\noi{\noindent}
\newcommand\seq{\;\;=\;\;}
\newcommand\barcaps{\cal}
\newcommand\jump{\vspace*{17pt}}
\newcommand\emptypage{~~~ \eject}
\setlength{\baselineskip}{17pt}
\def\be{\begin{eqnarray}}
\def\ee{\end{eqnarray}}
\def\pp{{+\hspace{-0.04in}+}}

\newenvironment{draftequation}[1]{\be\label{#1}}{\ee}
\newcommand\bbe[1]{\begin{draftequation}{#1}}
\newcommand\eee{\end{draftequation}}

\def\ps{p\hspace{-0.075in}/}     
\def\pis{\pi\hspace{-0.075in}/}  
\def\DD{\hbox{D\hspace{-0.10in}I\hspace{0.04in}}} 
\def\half{{\textstyle{1 \over 2}}}
\def\ihalf{{\textstyle{i \over 2}}}
\def\D{{\cal D}}
\def\L{{\cal L}}
\def\z{{\zeta}}
\def\l{{\lambda}}
\def\bl{{\bar\lambda}}
\def\del{\Delta}
\def\nab{\nabla}
\def\bna{{\bar\nabla}}
\def\d{\partial}
\def\ey{{e^{iY}}}
\def\bfi{{\bar\Phi}}
\def\ba{{\bar a}}
\def\bb{{\bar b}}
\def\fat{{\hat\Phi}}
\def\f{\Phi}
\def\bw{{\bar W}}
\def\bwl{{\bar w}}
\def\eab{\varepsilon^{ab}}
\def\eeab{\varepsilon_{ab}}
\newcommand\lra[1]{\mathord{\buildrel{
   \lower3pt\hbox{$\scriptscriptstyle\leftrightarrow$}}\over{#1}}}
%
\def\bop#1{\setbox0=\hbox{$#1M$}\mkern1.5mu
        \vbox{\hrule height0pt depth.04\ht0
        \hbox{\vrule width.04\ht0 height.9\ht0 \kern.9\ht0
        \vrule width.04\ht0}\hrule height.04\ht0}\mkern1.5mu}
\def\Box{{\mathpalette\bop{}}}

\def\c{\ , \ }
\def\q{\qquad\qquad\qquad}

\begin{titlepage}
\begin{flushright}
ITP-SB-94-23\\
USITP-94-10\\
hep-th/9406062\\
May 1994
\end{flushright}
\bigskip
\Large
\begin{center}
\bf{The Nonlinear Multiplet Revisited}\\

\bigskip

\normalsize
U.\ Lindstr\"om\footnote{email:ul@vana.physto.se} \\
{\it Institute of Theoretical Physics\\
University of Stockholm\\
Vanadisv\"agen 9\\
S-113 46 Stockholm SWEDEN}\\

\bigskip

Byungbae Kim\footnote{email:bkim@insti.physics.sunysb.edu}
and  M.\ Ro\v cek\footnote{email:rocek@insti.physics.sunysb.edu}\\
{\it Institute for Theoretical Physics\\
State University of New York\\
 Stony Brook, NY 11794-3840 USA\\}

\end{center}
\vspace{1.0cm}
\begin{abstract}
Using a reformulation of the nonlinear multiplet as a gauge
multiplet, we discuss its dynamics.  We show that the nonlinear
``duality'' that appears to relate the model to a conventional
$\sigma$-model introduces a new sector into the theory.
\end{abstract}
\end{titlepage}
\eject

\begin{flushleft}
\section{Introduction}
\end{flushleft}
\bigskip
Conformal field theories provide string backgrounds.  Different
supermultiplets in general give rise to different superconformal field
theories. The nonlinear multiplet was first introduced as a
compensating multiplet for $N=2$ $D=4$ supergravity \cite{dewit}.  It
was later used to construct hyperk\"ahler metrics \cite{KLR}. In this
letter we study it as a $D=2$ dynamical system in its own right. We
find a number of novel features: It can be formulated as a theory
with a sector linear in a gauge field analogous to a $\f F$
topological theory \cite{BF,witten}. It is ``dual'' to an ordinary
$\sigma$-model via a nonlinear ``duality'' that introduces new
solutions to the classical field equations (and presumably new states
in the quantum theory).

We begin this letter with a quick review of the nonlinear multiplet
and $N=4$ superspace. We then reformulate the nonlinear multiplet as
a kind of $U(1)$ gauge multiplet, and give a construction of a
nontrivial string theory background (a hyperk\"ahler manifold). Next
we define gauge-covariant components and give the component
action. Finally, we discuss the nonlinear ``duality''.

\bigskip

\begin{flushleft}
\section{Review of the nonlinear multiplet}
\end{flushleft}

\bigskip
We begin with a quick review of higher $N$ superspaces in two
dimensions. Dimensional analysis of the
superspace measure and the superfield component content implies that to
construct superspace actions one needs to find invariant subspaces and
corresponding restricted measures. Such subspaces are analogous to
$N=1$, $D=4$ chiral and antichiral superspaces. To this end, in a
series of papers \cite{GHR,LR,BLR,projective}, we have constructed and
used a projective superspace. In the present $N=4$ context, it is
introduced as follows \cite{KLR,LR} (see also \cite{GHR,BLR}): The
complex $SU(2)$ doublet spinor derivatives $D_{a\pm}$, $\bar D^b_{\pm}$
that describe $N=4$ supersymmetry obey the commutation relations
\bbe{CREL}
\left\{{D_{a\pm},\bar D^b_{\pm}}\right\}=i\delta
^b_a\d_{\buildrel \pp \over =}
\eee
(all others vanish). When we work with $N=2$ superfields, we
identify $D_\pm\equiv D_{1 \pm}$ as the $N=2$ spinor covariant
derivative and $Q_\pm \equiv D_{2\pm}$ as the generator of the
nonmanifest supersymmetries. We use a complex variable $\z$ to
define a set of anticommuting spinor derivatives\footnote{In
\cite{KLR,LR}, $\bna_\pm$ are rescaled by a factor:
$\bna_\pm\to-\z\bna_\pm$.}:
\bbe{NABLA}
\nab_\pm=D_\pm+\z Q_\pm\c \qquad \bna_\pm=\bar
D_\pm-\z^{-1}\bar Q_\pm\ .
\eee
A real structure $R$ acts on $\z$ by hermitian
conjugation composed with the antipodal map, i.e.;
\bbe{REAL}
R\z =-\bar\z^{-1}\ .
\eee
Since $\nab_\pm =R\bna_\pm$, the map $R$ preserves the
subspaces annihilated by the derivatives (\ref{NABLA}). To describe the
nonlinear multiplet \cite{dewit,KLR}, we consider a superfield
$\eta(\z)$, with particular $\z$ dependence and a reality condition:
\bbe{ETADEF}
\eta = \frac{a+b\z}{1+c\z}\c\qquad R\bar\eta=-\,\frac1\eta\ .
\eee
The reality condition implies that the $N=4$ superfields $a,b,c$
obey
\bbe{RCOND}
\bar a=-\,\frac{c}b\c\qquad \bar b=\frac1b
\eee
which we solve by writing
\bbe{etafi}
\eta=\frac{\bar\f+e^{iY}\z}{1-\f e^{iY}\z}\ .
\eee
Finally, we require that $\eta$ is
annihilated by the derivatives in (\ref{NABLA}); these $N=4$
constraints lead to the $N=2$ component relations
\bbe{DETA}
D_\pm\bfi=0&\c \qquad & Q_\pm\f=0\c \cr & &\cr
Q_\pm\bfi=-D_\pm [(1+\f\bfi)e^{iY}]&\c\qquad&
Q_\pm[(1+\f\bfi)e^{-iY}]= D_\pm\f\ ,
\eee
which imply
\bbe{ddeta}
D^2 [(1+\f\bfi)e^{iY}]=0\c
\eee
where $D^2\equiv D_+ D_-$.

To construct $N=4$ actions for $N=4$ superfields we use a second set of
linearily independent covariant spinor derivatives:
\bbe{DELTA}
\Delta_\pm=D_\pm-\z Q_\pm\c \qquad \bar\Delta_\pm=\bar
D_\pm+\z^{-1}\bar Q_\pm\ .
\eee
An action may then be written as
\bbe{N4ACT}
S=\frac1{16}\int{d^2x}\int_{C}{d\z} \Delta^2
\bar\Delta^2 \, L\left({\eta(\z );\z}\right)
\eee
where $C$ is an appropriate contour. Using
\bbe{delnab}
\del_\pm=2D_\pm-\nab_\pm \c\qquad \bar\del_\pm=2\bar
D_\pm-\bar\nab_\pm\c
\eee
and (of course) $\nab_\pm\eta=\bar\nab_\pm\eta=0$, the $N=2$ superspace
form of the
action (\ref{N4ACT}) is:
\bbe{N2ACT}
S=\int{d^2x}D^2\bar D^2\int_{C}{d\z}L\left({\eta(\z);\z
}\right)
\eee
Up to a sign ($\eta\to -\eta$) and the relabelling
($X\to Y$, $\chi\to\f$), this is the description of
the nonlinear multiplet given in \cite{KLR}.

We may also consider an $N=2$ nonlinear multiplet that obeys only the
constraint (\ref{ddeta}) and
\bbe{chir}
\bar D_\pm \f=0\ ,
\eee
and has an action
\bbe{n2act2}
S=\int{d^2x}D^2\bar D^2L\left(\f ,\bfi ,Y\right)
\eee
for arbitrary $L$. Clearly, the $N=4$ symmetric action (\ref{N2ACT})
is a special case of this, and our subsequent discussion applies to
the general case (\ref{n2act2}).

We do not know how to solve the constraints
(\ref{ddeta}) and (\ref{chir}) in superspace (Actually, not even in
$N=1$ superspace!). Consequently, it is not clear what the superfield
equations are, and in \cite{KLR} we did not look directly at the
dynamics of the nonlinear multiplet. To do so, we would have needed to
go to components, and the component expansion of the action
(\ref{n2act2}) subject to the constraints (\ref{ddeta}), (\ref{chir})
is very tedious. In the next section, we reformulate the nonlinear
multiplet in a way that makes the computation of the component action
tractable.

In \cite{KLR}, we studied the action in another way: We found a
``duality'' transformation to a formulation in terms of ordinary
$N=2$ chiral superfields; in terms of these, the dynamics can be
understood straightforwardly. The ``duality'' transformation
resembles the well known target space duality of string theory, but
is nonlinear.  It is performed as follows: we replace the action by a
first-order action
\bbe{n2act1}
S_1=\int{d^2x}D^2\bar D^2\left( L(\f ,\bfi ,\Psi ) +(1+\f\bfi
)(\chi e^{-i\Psi}+\bar\chi e^{i\Psi})\right)\ ,
\eee
where $Y\to\Psi$ is now an unconstrained superfield, and $\chi$ is a
chiral superfield: $\bar D_\pm \chi = 0$. Integrating out $\chi$, we
recover the constraint (\ref{ddeta}); integrating out $\Psi$ gives
$\Psi (\f ,\bfi ,\chi ,\bar\chi)$, and substituting back gives a
standard $N=2$ action for the chiral superfields $\f ,\chi$.

However, as noted in \cite{KLR}, this is a peculiar ``duality'': the
trivial action $S=0$ gives rise to a {\em non}trivial free action
for $\f ,\chi$.  We resolve this paradox in the section 4.

\bigskip

\begin{flushleft}
\section{The nonlinear multiplet as a gauge multiplet.}
\end{flushleft}
As discussed in the introduction, the formalism can be simplified by
rewriting the nonlinear multiplet as a gauge multiplet.  We do this
by rescaling
\bbe{rescale}
\f\to\frac{\f^1}{\f^2}\c\qquad
e^{iY}\to e^{iY}\frac{\bfi_2}{\f^2}\ .
\eee
Then the superfield $\eta$ becomes
\bbe{etagauge}
\eta=\frac{\bfi_1+\ey\f^2\z}{\bfi_2-\f^1\ey\z}\ ,
\eee
and the constraints and transformations (\ref{DETA}) become
(suppressing the $\pm$ indices on the spinor operators $D,Q$):
\bbe{detag}
\eab\bfi_aD\bfi_b = 0 &\c\ \ & \eeab\f^aQ\f^b=0\c\cr &&\cr
\eab\bfi_aQ\bfi_b=-(\f^a\ey)\lra{D}\bfi_a&\c\ \ &
\f^a\lra{Q}(e^{-iY}\bfi_a)=-\eeab\f^aD\f^b\ .
\eee
The superfield $\eta$ is left unchanged by a gauge transformation:
\bbe{gaugetrans}
\f^a\to e^\Lambda \f^a\c\qquad Y\to Y+i(\Lambda-\bar\Lambda)\ .
\eee
Because of this invariance, the transformations (\ref{detag}) do not
determine the transformations of the fields $\f^a$; We can
consistently choose:
\bbe{fixtrans}
Q\f^a=0\c\qquad D\bfi_a=0\c\qquad Q(|\f
|^2e^{-iY})=-\eeab\f^aD\f^b
\eee
and
\bbe{fixtranstoo}
Q\bfi_a=-\eab\frac{\bfi_b}{\bfi_c\bfi_c}D(|\f |^2e^{iY})\ .
\eee
(Note that (\ref{fixtranstoo}) breaks the manifest $SU(2)$ invariance
of (\ref{detag}) to an $SO(2)$ subgroup.)  The transformations and
constraints (\ref{fixtrans}), (\ref{fixtranstoo}) are invariant under
the gauge transformations (\ref{gaugetrans}) as long as $\Lambda$ is
chiral: $\bar D_\pm\Lambda=0$. In this case, $Y$ transforms as an
ordinary $N=2$ $U(1)$ gauge supermultiplet. This allows us to find
component actions in a Wess-Zumino gauge, greatly simplifying our
calculations.

We work in a chiral representation of the gauge group, and define
\bbe{fihat}
\hat\f^a\equiv \bfi_ae^{-iY}\ \Rightarrow \ \eta =
\frac{\fat^1+\f^2\z}{\fat^2-\f^1\z}\ .
\eee
We also define gauge covariant derivatives $\nab$
\bbe{gnab}
\nab = e^{-iqY}De^{iqY}\c\qquad \bna = \bar D\c
\eee
where $q$ is the charge of the field that $\nab$ acts on ($\f$ and
$\fat$ both have $q=1$). These derivatives obey the usual algebra
\bbe{aljabric}
\{\nab_+,\nab_-\}=0&\c \qquad &\{\bna_+,\bna_-\}=0\c \cr
&&\cr
\{\nab_+,\bna_+\}=\nab_\pp&\c \qquad &\{\bna_-,\nab_-\}=\nab_=\c \cr
&&\cr
\{\nab_+,\bna_-\}=-\bar Wq&\c \qquad &\{\bna_+,\nab_-\}=Wq\c \cr
&&\cr
[\nab_+,\nab_=]=(\nab_-\bar W)q&\c \qquad&
[\bna_+,\nab_=]=-(\bna_- W)q\c\cr
&&\cr
[\nab_-,\nab_\pp ]=-(\nab_+ W)q&\c \qquad&
[\bna_-,\nab_\pp ]=(\bna_+ \bar W)q\c\cr
&&\cr
[\nab_\pp ,\nab_=]=fq &\c \qquad& f\equiv \bna_+\nab_-\bar W
-\nab_+\bna_- W\ .
\eee
Here $W=i\bar D_+D_-Y$ is the superfield strength of $Y$, and is a
twisted chiral superfield with charge $q=0$: $\bar D_+W=D_- W=0$.

The $N=2$ action (\ref{n2act2}) becomes:
\bbe{n2actg}
S=\int{d^2x}D^2\bar D^2\,L(\f^a ,\fat^a)\ ,
\eee
where $L$ is restricted to be gauge invariant\footnote{Actually, gauge
invariance of the action implies invariance of $L$ only up to
superspace total derivatives; this has no effect on our analysis.}:
\bbe{linv}
\f^aL_a+\fat^aL_\ba=0\c
\eee
and where $L_a\equiv\frac{\d L}{\d\f^a}$ and $L_\ba\equiv\frac{\d
L}{\d\fat^aa}$. We use (\ref{linv}) and its derivatives frequently
below. The condition that the action (\ref{n2actg}) has $N=4$
supersymmetry is Laplace's equation
\bbe{n4L}
L_{a\ba}=0\ .
\eee

The constraint (\ref{ddeta}) (and its complex conjugate) become
simply
\bbe{ddabs}
\nab^2(\f^a\fat^a) = \bna^2(\f^a\fat^a) =0 \ .
\eee
These constraints are
actually gauge {\em covariant}; thus when we go to first order form
and impose them via chiral Lagrange multipliers $\chi$ and
$\hat\chi\equiv\bar\chi e^{2i\Psi}$ (recall that we are replacing
$Y$ with the unconstrained gauge superfield $\Psi$), the action
(\ref{s1g}) is gauge invariant if $\chi ,\hat\chi$ both have charge
$q=-2$:
\bbe{s1g}
S_1=\int{d^2x}D^2\bar D^2\,\left[ L(\f^a ,\fat^a ) +(\chi+\hat\chi
)\f^a\fat^a \right] \ .
\eee
The gauge invariance of the action allows us to choose a gauge that
is very convenient for performing the ``duality'' transformation:
$\chi=\half$.  In this gauge, we may for example consider $L=i\,
ln\left(\frac{\f^1}{\fat^1}\right)$; this gives
\bbe{exam}
S_1=\int{d^2x}D^2\bar D^2\,\left[\Psi + |\f |^2 cos(\Psi)\right]\ .
\eee
(Equivalently, we may choose as the action a Fayet-Iliopoulos term for
$\Psi$.) Eliminating $\Psi$ by the equation $sin(\Psi )= |\f
|^{-2}$, we get the K\"ahler potential for the Eguchi-Hansen
gravitational instanton with the ``wrong'' sign of the mass-parameter.
This gives a nontrivial example of a hyperk\"ahler metric constructed
using the nonlinear multiplet and ``duality'' transformation.
\bigskip

\begin{flushleft}
\section{Components}
\end{flushleft}

In this section, we descend from superspace to spacetime, and
compute the component action.  We work with the first order
system (\ref{s1g}), and derive the component form of the
constraints by integrating out the component Lagrange multiplier
fields. After giving the full action in a compact geometric
formulation, we focus on the bosonic sector and resolve the paradox
that our nonlinear ``duality'' introduced.

As usual, we define component fields as $\theta$ independent
projections of the superfields and their spinor derivatives. The
Wess-Zumino gauge components of the gauge superfield are:
\bbe{gfcomp}
&\nab_\pp|=\d_\pp +V_\pp\c\qquad \nab_=|=\d_=
+V_=\c\qquad W|=w\c\qquad \bw |=\bar w\c &\cr &&\cr
&\nab_+W|=\l_+\c\qquad\bna_-W|=\l_-\c\qquad\bna_+\bw
|=\bl_+\c\qquad\nab_-\bw |=\bl_-\c&\cr &&\cr
&(\bna_+\nab_-\bw-\nab_+\bna_-W)|=f\equiv\d_\pp V_= -
\d_=V_\pp\c&\cr&&\cr
& (\bna_+\nab_-\bw+\nab_+\bna_-W)|=i\DD\c&
\eee
where $V$ is the component gauge field, and $w,\l,\DD$, \etc, are
various gauge invariant superpartners of $V$.

The components of the chiral superfields $\f^a,\chi$
(recall that $\chi$ is the Lagrange multiplier field) are:
\bbe{fccomp}
&\f^a|=A^a\c\qquad\nab_\pm\f^a|=\psi_\pm^a\c\qquad
i\nab_+\nab_-\f^a|= F^a\c&\cr&&\cr
&\chi |=C\c\qquad\nab_\pm\chi |=\chi_\pm \c\qquad
i\nab_+\nab_-\chi |=G\ .&
\eee
We denote the set $(\f^a,\chi)$ collectively by $\f^i$ (similarly, we
denote $A^i=(A^a,C)$, \etc), and the total super-Lagrangian
$\left( L(\f^a ,\fat^a ) +(\chi+\hat\chi )|\f |^2\right)$ by the
K\"ahler potential $K(\f^i ,\fat^i)$.

The component Lagrangain, after integrating by parts and eliminating
the auxiliary fields $F^i$ to bring out the geometric features, is:
\bbe{compact}
\L=&G_{i\bar j}\left( \half(\nab_\pp A^i\nab_=\bar A^j+\nab_=
A^i\nab_\pp \bar A^j)+w\bwl k^i\bar k^j+w\psi^i_+\bar
k^j_{;l}\bar\psi^l_--\bwl\bar\psi^j_+k^i_{;l}\psi^l_-\right. &\cr&&\cr
&\left.\qquad-\psi^i_+\D_=\bar\psi^j_+
-\psi^i_-\D_\pp\bar\psi^j_-+\bar\l_+\bar k^j\psi^i_-+\psi^i_+
\bar k^j\l_- -\l_+k^i\bar\psi^j_--\bar\psi^j_+
k^i\bar\l_-\right)&\cr&&\cr
&+R_{i\bar jk\bar l}\psi^i_+\psi^k_-
\bar\psi^j_+\bar\psi^l_-+{\textstyle{i\over4}}(\bar k^iK_{\bar
i}-k^iK_i)\DD\  ,\q\qquad\qquad\qquad\quad&
\eee
where the metric $G_{i\bar j}$, the
Levi-Civita connection $\Gamma^i_{jk}$, and the curvature $R_{i\bar
jk\bar l}$ of the K\"ahler manifold with potential $K(A^i,\bar
A^i)$ are
\bbe{gij}
&&\cr
&G_{i\bar j}=K_{i\bar j}=\left( {\matrix{{L_{a\bar b}+\delta
_{a\bar b}(C +\bar C)}&{\bar A^a}\cr &\cr {A^b}&0\cr
}} \right)\c&\cr &&\cr\cr &&\cr
&\Gamma^i_{jk}=(G_{i\bar l})^{-1}K_{\bar l jk}\c&\cr &&\cr
&R_{i\bar jk\bar l}=K_{ik\bar j\bar l}-(G_{m\bar n})^{-1}K_{m\bar j
\bar l}K_{\bar n ik}\c&
\eee
(for K\"ahler manifolds, these and their complex conjugates are the
only nonvanishing components). Further, $k^i$ is the killing vector
$k^i=(A^a,-2C)$, $k^i_{;j}=\d_i k^i+\Gamma^i_{jl} k^l$, and the
world-sheet covariant derivatives
\bbe{cald}
\nab A^i&=& \d A^i +V k^i\c\cr&&\cr
\D \bar\psi^i&=&\nab\bar\psi^i+\Gamma^{\bar i}_{\bar j\bar k}
\nab\bar A^j\bar\psi^k\cr&&\cr
&=&\d\bar\psi^i+V\bar k^i_{;j}\bar\psi^i
+\Gamma^{\bar i}_{\bar j\bar k}\d\bar A^j\bar\psi^k
\eee
are both gauge and diffeomorphism covariant. In deriving the
Lagrangian $\L$ (\ref{compact}), we have used the following identities:
\bbe{ident}
\nab_+\bna^2\fat^i&=&\left(\nab_\pp\bna_-+(\bna_+\bw )q+\bw
q\bna_+\right)\fat^i\c\cr&&\cr
\nab_-\bna^2\fat^i&=&\left(-\nab_=\bna_++(\bna_-W)q+W
q\bna_-\right)\fat^i\c\cr&&\cr
\nab^2\bna^2\fat^i&=&\left(-\half\Box +\half
(\bna_+\nab_-\bw+\nab_+\bna_-W)q\right.\cr&&\cr
& &\left.\ -W\bw q^2 -\nab_-\bw q\bna_++\nab_+W
q\bna_-\right)\fat^i\ ,
\eee
where $q$ is the charge of $\fat^i$. To get the pure
nonlinear multiplet action, we separate out the dependence on the
Lagrange-multiplier multiplet
$\chi$, and integrate it out. This gives the constraints:
\bbe{compcon}
&&A^a\bar F^a = 0 \c \qquad \psi^a_+\bar F^a +iA^a(\nab_\pp
\bar\psi^a_-+\bl_+\bar A^a+\bwl\bar\psi^a_+) = 0 \c \cr&&\cr
&&\psi^a_-\bar F^a +iA^a(-\nab_= \bar\psi^a_++\l_-\bar
A^a+w\bar\psi^a_-) = 0 \c\cr &&\cr
&&F^a\bar F^a -\psi^a_+(-\nab_= \bar\psi^a_++\l_-\bar
A^a+w\bar\psi^a_-)+\psi^a_-(\nab_\pp \bar\psi^a_-+\bl_+\bar
A^a+\bwl\bar\psi^a_+)\cr &&\cr
&&\qquad\qquad-A^a(-\half\Box\bar A^a+(\ihalf\DD -w\bwl)\bar A^a
-\bl_-\bar\psi_+^a +\l_+\bar\psi_-^a)=0\ ,
\eee
and their complex conjugates.

The basic features of the model can be seen in the bosonic
sector of the Lagrangian (\ref{compact}); after integrating by parts
and collecting terms, we find
\bbe{boscomp}
\L_{bose}=&L_{a\bar b}\left( A^a\bar A^bw\bwl
+\half (\nab_\pp A^a\nab_=\bar A^b+\nab_=A^a\nab_\pp\bar
A^b\right) \qquad\qquad\qquad\qquad&\cr&&\cr
&+{\textstyle{i\over4}}(\bar A^a L_{\bar
a}-A^aL_a)\DD-(C+\bar C)\left( w\bwl |A|^2+{\textstyle{1\over4}} (A^a
{\Box} \bar A^a+\bar A^a \Box A^a) \right) &\cr&\cr
&-\ihalf (\bar C- C)(\DD |A|^2-\ihalf\bar A^a
\lra{{\Box}}A^a)\ ,\q\q\qquad&
\eee
where $\Box \equiv \{\nab_\pp ,\nab_=\}$. This Lagrangian
is gauge invariant; one may pick various gauges, \eg,
$|A^1|=1$ or $|C|=1$. Integrating out the gauge multiplet fields
$w,V,\DD$ gives the ``dual'' theory: the $w$ field equation sets
$\bar w=0$, the $V$ field equation gives $V(A,C)$, and
the $\DD$ field equation gives $\bar C- C$ in terms of $A$. After
substituting back into the Lagrangian $\L_{bose}$, in the gauge $|C|=1$
this is just an ordinary $\sigma$-model with target space coordinates
$A^a,\bar A^a$; the actual form of the resulting $\L_{dual}$ is most
easily found by going back to the superspace action (\ref{s1g}) in
the gauge $\chi=\half$ and integrating out $\Psi$ to find the K\"ahler
potential.

Integrating out the Lagrange multiplier field $C$ gives
\bbe{constr}
\DD= \frac{i\bar A^a\lra{{\Box}}A^a}{2|A|^2}\c\qquad
w\bar w=-\frac{A^a{\Box} \bar A^a+\bar A^a \Box A^a}{4|A|^2}\ .
\eee
Substituting back into $\L_{bose}$ (\ref{boscomp}), we find
\bbe{weird}
&\L_{nonlin}=L_{a\bar b}\left( -A^a\bar A^b\frac{A^c{\Box} \bar
A^c+\bar A^c \Box A^c}{4|A|^2} +\half (\nab_\pp A^a\nab_=\bar
A^b+\nab_=A^a\nab_\pp\bar A^b)\right) &\cr&&\cr
&-{\textstyle{1\over8}}(\bar A^a L_{\bar
a}-A^aL_a)\frac{\bar A^c\lra{{\Box}}A^c}{|A|^2}\ .\q\qquad\qquad&
\eee
This still depends on the gauge field $V$; however, the dependence is
now only {\em linear}, and hence $V$ cannot be integrated out.  Thus
we appear to have found a ``duality'' between a conventional
$\sigma$-model and a model with a sector that is first order in
derivatives. The most extreme form of this paradox occurs when the
superspace lagrangian $L$ (\ref{n2actg}) vanishes: then
$\L_{nonlin}=0$, but its ``dual'' is $\L_{dual}=-\half A^a\Box\bar
A^a$. To understand this peculiar result, we must examine how we
integrate out $w$ more carefully. Varying $w$ in $\L_{bose}$
(\ref{boscomp}) gives
\bbe{wfeq}
\bar w \left[ L_{a\bb}A^a\bar A^b-(C+\bar C)|A|^2\right]=0\ .
\eee
This has {\em two} solutions: The one we naively took --- which
one normally takes when eliminating auxiliary fields --- namely, $\bar
w=0$, and $C+\bar C=\frac{L_{a\bb}A^a\bar A^b}{|A|^2}$.  If we take
the first solution $\bar w=0$, then we get the ordinary
$\sigma$-model; however, the second solution $C+\bar
C=\frac{L_{a\bb}A^a\bar A^b}{|A|^2}$ gives precisely the peculiar
model with Lagrangian $\L_{nonlin}$. Thus the two models are actually
two different sectors of the first order theory described by
$\L_{bose}$ (\ref{boscomp}). In sum: the first order Lagrangian has
two sectors.  One, which is reached by integrating out the Lagrange
multiplier multiplet, is the nonlinear multiplet model.  When one
instead attempts to integrate out the gauge multiplet from the first
order action, one finds that depending on the order of integration and
a choice of solutions to a nonlinear algebraic equation, one gets
either the original nonlinear multiplet model (sector one), or a new
sector corresponding to an ordinary
$\sigma$-model.

\bigskip

\begin{flushleft}
\section{Conclusions}
\end{flushleft}

We have reformulated the nonlinear multiplet as a gauge multiplet.
This allowed us to compute the component Lagrangian in a Wess-Zumino
gauge. We found a system with a subsystem linear in the gauge field
(\ref{weird}). We considered a nonlinear ``duality'' transformation
that eliminates this unusual subsystem and gives an ordinary
$\sigma$-model.  We found that the two theories are not dual in the
sense of representing different formulations of the same theory, but
correspond to different sectors of the first order model
(\ref{boscomp}).

Our analysis has been entirely classical; it would be interesting to
study the nonlinear multiplet at the quantum level. In particular,
one would like to see if the nonlinear ``dual'' of a superconformal
field theory is superconformal, and if it is, what, if any, relation
the two theories have. A second unsolved problem is classical and
more geometric: which $\sigma$-models admit a nonlinear ``duality''
to a nonlinear multiplet formulation.

\bigskip
\begin{flushleft}
{\bf Acknowledgments}
\end{flushleft}

It is a pleasure to thank the ITPs at Stony Brook and Stockholm,
as well as the Physics Department at Oslo University, for hospitality.
We would like to thank Jan de Boer for his comments on the manuscript.
UL acknowledges partial support from the NFR under Grant No.\ F-FU
4038-300 and NorfA under Grant No.\ 93.35.088/00, and MR
and BBK acknowledge partial support from the NFS under Grant No.\
PHY 93 09888.

\end{document}